\documentclass[letterpaper,twoside,12pt]{article}
\usepackage{amssymb}
\usepackage{amsmath}
\usepackage{latexsym}
\usepackage{verbatim}
\usepackage{graphicx}
\usepackage{epsfig}
\usepackage{stmaryrd}
\usepackage{rotating,graphicx}
\usepackage[english]{babel}
\usepackage{setspace}
\usepackage[english]{babel}
\usepackage[utf8]{inputenc}
\usepackage[T1]{fontenc}
\usepackage{lmodern}

\begin{document}

\newcommand{\bra}[1]{\langle #1|}
\newcommand{\ket}[1]{|#1\rangle}
\newcommand{\braket}[2]{\langle #1|#2\rangle}

\begin{Large}
\begin{center}
\textbf{The Consistent Histories Formalism and the Measurement Problem}\\
\end{center}
\end{Large}

\begin{center}
\begin{large}
Elias Okon\\
\end{large}
\textit{Instituto de Investigaciones Filos\'oficas, Universidad Nacional Aut\'onoma de M\'exico, Mexico City, Mexico.\\ E-mail:} \texttt{eokon@filosoficas.unam.mx}\\[.5cm]
\begin{large}
Daniel Sudarsky\\
\end{large}
\textit{Instituto de Ciencias Nucleares, Universidad Nacional Aut\'onoma de M\'exico, Mexico City, Mexico.\\ E-mail:} \texttt{sudarsky@nucleares.unam.mx}\\[.5cm]
\end{center}

\noindent \textbf{Abstract:} In response to a recent rebuttal of \cite{Oko.Sud:14b} presented in \cite{Gri:15}, we defend the claim that the Consistent Histories formulation of quantum mechanics does not solve the measurement problem. In order to do so, we argue that satisfactory solutions to the problem must not only not contain anthropomorphic terms (such as \emph{measurement} or \emph{observer}) at the fundamental level, but also that applications of the formalism to concrete situations (e.g., measurements) should not require any input not contained in the description of the situation at hand at the fundamental level. Our assertion is that the Consistent Histories formalism does not meet the second criterion. We also argue that the so-called \emph{second} measurement problem, i.e., the inability to explain how an experimental result is related to a property possessed by the measured system \emph{before} the measurement took place, is only a \emph{pseudo-problem}. As a result, we reject the claim, defended in \cite{Gri:15}, that the capacity of the Consistent Histories formalism to solve it should count as an advantage over other interpretations.

\section{Introduction}
\label{Int}
\onehalfspacing
The Consistent Histories (CH) framework provides a formulation of quantum mechanics that assigns probabilities to histories of all kinds of systems, microscopic and macroscopic, using a single universal machinery and without ``Heisenberg cuts'' or references to measurements or observers. As a result, proponents of CH maintain that the formalism overcomes the measurement problem of the standard 
interpretation (as well as \emph{all} other standard quantum paradoxes). In \cite{Oko.Sud:14b} we 
have disputed such an assertion by displaying an array of conceptual problems with the way the 
formalism is deployed in measurement situations.\footnote{Other objections against CH can be found in \cite{Esp:87,Dow.Ken:96,Ken:97,Bar:99,Ken:10,Oko.Sud:14a}.} In \cite{Gri:15}, however, arguments 
against our objections are presented, and so the main objective of this article is to respond to such a 
challenge. We hope that, by doing so, we will not only be able to adequately defend our position, but 
also to shed light on the root of the disagreement. In this regard, we believe that the origin of the 
dispute arises from a difference on what CH proponents and us take the measurement problem to be, 
and, more importantly, on what CH proponents and us regard as a \emph{satisfactory solution to the 
problem}. In short, we believe that such a solution must not only avoid making any reference 
to \emph{measurements} or \emph{cuts} at the fundamental level, but also that successful applications of 
the formalism must not depend on input not present in the fundamental theory. Our claim, in a nutshell, 
is that CH accomplishes the first but not the second. 

In the rest of the paper we develop these ideas. To do so, we briefly review the CH formalism in section \ref{CH} and in section \ref{MP} we discuss what it takes to solve the measurement problem. Then, in section \ref{MCH} we summarize our arguments in \cite{Oko.Sud:14b} and in section \ref{Rep} we evaluate and respond to the challenges raised in \cite{Gri:15}. Finally, in section \ref{C} we present our conclusions.

\section{A brief presentation of the Consistent Histories formalism}
\label{CH}
Before getting down to business, we will briefly describe the CH formalism (see \cite{Gri:03} for a comprehensive presentation). As we said above, CH assigns probabilities for all systems, microscopic or macroscopic, using the same machinery and without any reference to measurements or cuts. More specifically, the most general objective of CH is the prediction of probabilities for time histories of systems, where histories are defined as sequences of properties and are represented by projection operators at successive times. CH, then, introduces the notion of sets of histories and specifies rules that assign probabilities to the various elements of each set. However, according to CH, not all sets of histories allow for probabilities to be assigned. This is possible only when: i) the sum of probabilities of all members of a set equals one, and ii) all pairs of histories within the set are orthogonal. Families satisfying these two conditions are called \emph{frameworks}, or \emph{realms}. 

A natural consequence of the CH formalism is that, given a system, multiple incompatible frameworks can be constructed (i.e., different frameworks that assign incompatible properties to the system). Therefore, in order to avoid inconsistencies, CH requires the imposition of the following rules:
\begin{itemize}
\item \textbf{Single-Framework Rule}: probabilistic reasoning is invalid unless it is carried out using a single framework.
\item \textbf{Principle of Liberty}: one can use whatever framework one chooses in order to describe a system.
\item \textbf{Principle of Equality}: all frameworks are equally acceptable in terms of fundamental quantum mechanics.
\item \textbf{Principle of Utility}: not all frameworks are equally useful in answering particular questions of physical interest.
\end{itemize}
This, however, comes with a price because the enforcement of these rules leads to the violation of the following principle:
\begin{itemize}
\item \textbf{Principle of Unicity}: alternative descriptions of physical systems always can be combined into a single unified one, from which all views can be derived as partial characterizations.
\end{itemize}
Whether this is too high a price to pay is an interesting question. However, what we would like to point out for now is that, as we will see in section \ref{Rep}, and contrary to what is claimed in \cite{Gri:15}, none of the objections that were presented in \cite{Oko.Sud:14b} are based on the fact that the Principle of Unicity is not valid within CH.


\section{Solving the measurement problem}
\label{MP}
A lot has been written about the measurement problem of quantum mechanics. A popular way to describe it, among many, is the following: even though quantum mechanics depends crucially on the notion of measurement, such notion is never formally defined within the theory. Then, in order to use quantum mechanics, one needs to know, \emph{by means external to the theory}, what constitutes a measurement. Of course, the measurement problem is a problem of a theoretical framework and so, in order to state it, one needs to first specify in detail the theoretical framework in question.\footnote{The formulations of the measurement problem developed in \cite{Mau:95}, instead of specifying in detail the theoretical framework to be dealt with, imposes general restrictions that all satisfactory formulations must obey.} This, given the proliferation of views regarding quantum mechanics, leads to a proliferation of ways to state the problem. For example, the description given above is suitable for Dirac's or von Neumann's formulation but does not apply to Bohr's, where the problem manifests as an ambiguity regarding where the classical-quantum cut should be drawn. It also does not apply to a formulation with a purely unitary evolution, where the problem manifests as a mismatch between experience and some predictions of the theory. 

At any rate, for the purposes of this paper it is sufficient to note that both us, and the author of \cite{Gri:15}, agree on the fact that the standard or orthodox interpretation suffers from the measurement problem (see e.g. \cite[p. 214]{Gri:03}). What we consider more important, given the objective of this work (i.e., evaluating whether CH solves or not the measurement problem), is a discussion of what constitutes a valid solution to the problem. In this regard, \cite{Gri:15} offers the following:
 \begin{quotation}
\noindent If quantum mechanics applies not only to the microscopic world of nuclei and atoms, but also to macroscopic objects and things that are even larger - from the quarks to the quasars - then the measurement process in which an earlier microscopic property is revealed in a macroscopic outcome should itself be describable, at least in principle, in fully quantum mechanical terms. Applied equally to the system being measured and to the macroscopic apparatus, and without the evasion and equivocation ridiculed by Bell \cite{Bel:90}. (\cite[p. 3]{Gri:15})
 \end{quotation}
Namely, if quantum mechanics applies to everything - from quarks to quasars - then measurements must be fully describable in purely quantum terms. Of course, one could hold that quantum mechanics does not apply to everything, but in such a case one would need to clearly establish where to draw the line (i.e., where to insert the ``Heisenberg cut'') - something that no one has been able to achieve. In any case, the CH formalism assumes that quantum mechanics does apply to everything so we will stick to such a premise. The quote also mentions that the application of the quantum formalism to measurement scenarios must not involve ``the evasion and equivocation ridiculed by Bell'' in \cite{Bel:90}. So what does Bell say in \cite{Bel:90} regarding a satisfactory quantum formalism (i.e., one that solves the measurement problem)? He concisely states the following:
 \begin{quotation}
\noindent The theory should be fully formulated in mathematical terms, with nothing left to the discretion of the theoretical physicist. (\cite[p. 33]{Bel:90})
 \end{quotation}
Then, according to Bell, there are two main components required by a valid solution for the measurement problem:
\begin{enumerate}
\item \textbf{The theory should be fully formulated in mathematical terms}: i.e., concepts such as \emph{measurement}, \emph{measuring apparatus}, \emph{observer} or \emph{macroscopic} should not be part of the fundamental language of the theory.
\item \textbf{Nothing should be left to the discretion of the theoretical physicist}: i.e., successful applications of the theory must not require any input not contained in the description of the situation at hand at the fundamental level.
\end{enumerate}
The point, then, is that in order to solve the measurement problem it is not enough to construct a formalism fully written in precise terms. One must also make sure that successful applications of the formalism do not require the introduction of information that is not already contained in the fundamental description given by the theory of the situation one wants to consider. That is, once a complete quantum description of the measurement scenario is given, including the quantum state of the apparatus and the full Hamiltonian (and remember that we are assuming that, at least in principle, that is always possible because we are assuming that quantum mechanics applies to everything), then, with that information alone, one must be able to use the theory to make concrete predictions regarding the possible final outcomes of the experiment.

It is important to emphasize that the above mentioned restriction to introduce ``further information not contained in the description at the fundamental level'' does not preclude the full specification of the physical situation characterizing the experiment. On the contrary, one is expected to provide the full fledged quantum state of the complete system (which consists of the sub-system of interest together with all the devices and apparatuses involved), as well as the full Hamiltonian characterizing the behavior of the sub-system, all devices and their interactions with the sub-system. It is clear that without such detailed characterization of the situation it is impossible to make concrete prediction. The important issue is whether a certain approach, provided with all the elements mentioned above, is capable or not to produce the specific predictions (even if these are probabilistic in nature) that correspond to what is found in actual experiments.

Let's illustrate this issue with a simple example, first from the perspective of the standard interpretation. It consists of a free spin-$\frac{1}{2}$ particle, with initial state 
\begin{equation}
\ket {\psi (0)} = \ket {+_x} = \frac{1}{\sqrt{2}} \left  (\ket {+_z}+\ket {-_z} \right ) ,
\end{equation}
to be measured by a suitable apparatus. The apparatus contains a macroscopic pointer, whose center of mass $y$ has an initial (ready) state given by a narrow wavefunction $\varphi (y)$ centered at $y=0$. 
For simplicity, we take the pointer's free Hamiltonian and the one corresponding to the spin degree of freedom to be zero. We also ignore the degree of freedom associated to the position of the particle.\footnote{None of these simplifying assumptions impinges on the conceptual validity of the example.} The spin and the apparatus interact via the interaction Hamiltonian:
\begin{equation}
\hat{H}_{I}=2 i \hbar \lambda \left( \frac{\partial}{\partial y} \right) \otimes S_z
\end{equation}
with $\lambda$ a constant. Given our assumptions, the interaction Hamiltonian also represents the complete Hamiltonian $\hat{H}$ of the whole system. Thus, putting everything together, we have that the initial state of the complete system is given by
\begin{equation}
\left\vert \Psi(0)\right\rangle = \varphi(y)\otimes \ket {\psi (0)} = \varphi(y) \otimes \frac{1}{\sqrt{2}} \left  (\ket {+_z}+\ket {-_z} \right  ) ,
\end{equation}
and that the total Hamiltonian is
\begin{equation}
\hat{H}=2 i \hbar \lambda \left( \frac{\partial}{\partial y} \right) \otimes S_z .
\end{equation}
The situation is thus, according to the theory, described in full. We have provided the entire Hamiltonian 
of the complete system, including 
the measuring apparatus, and the initial state of the whole system as well. 

The next step is to use the dynamical law of the theory (i.e., the Schrödinger equation), from which it is easy to show that the state of the complete system at time $t$ is given by
\begin{equation}
\ket {\Psi(t)} =e^{-\frac{i}{\hbar}\hat{H}t}\left\vert \Psi(0)\right\rangle = \frac{1}{\sqrt{2}}  \left [ \varphi(y-\lambda t) \otimes  \ket {+_z}+ \varphi(y+\lambda t) \otimes  (\ket {-_z} \right  ] .
\label{sup}
\end{equation}
Therefore, if $\lambda t$ is big enough, one can infer the value of the spin along $z$ by looking at the position of the pointer. 

Note however, that according to standard quantum mechanics, the final state can perfectly well be written as
\begin{equation}
\ket {\Psi(t)} = \frac{1}{\sqrt{2}}  \left [ \varphi_+(y,t) \otimes  \ket {+_x}+ \varphi_-(y,t) \otimes  (\ket {-_x} \right  ] 
\end{equation}
with
\begin{align}
\varphi_+(y,t)  &  \equiv \frac{1}{\sqrt{2}} \left[  \varphi(y-\lambda t) +\varphi(y+\lambda t) \right] , \nonumber\\
\varphi_-(y,t)  &  \equiv \frac{1}{\sqrt{2}} \left[  \varphi(y-\lambda t) - \varphi(y+\lambda t) \right] , \nonumber
\end{align}
so it seems that one could also use the system to measure the spin along $x$ by projecting the state of the center of mass of the pointer unto $\varphi_+(y,t)$ and $\varphi_-(y,t)$. We of course \emph{know} that if we perform the experiment in the laboratory we will end up with either $\varphi(y-\lambda t)$ or $\varphi(y+\lambda t)$ and not with either $\varphi_+(y,t)$ or $\varphi_-(y,t)$. The first two are perfectly sensible, well-localized states for a macroscopic object but the latter two represent bizarre ``Schrödinger cat'' states. The issue of course is how do we know this? Is the standard interpretation capable of \emph{predicting} it? Certainly not! The truth is that our knowledge of which one is the appropriate basis comes from the experience we have with macroscopic objects: we know that they always possess well-defined positions. The problem is that standard quantum mechanics is unable to \emph{account} for this because it is impossible to derive such a result from the standard formalism alone, even if as above, the complete system is completely described to the extent required by the theory. 

The natural question, then, is how does standard quantum mechanics manage to be as successful as it is, in spite of such a glaring deficiency? The answer is that, in order to make such accurate predictions, it requires to be \emph{implicitly} supplemented by external information regarding what it is that the apparatus one uses actually measures (the spin along $z$ in the above example, or, more generally, the fact that macroscopic objects always possess well-localized states). Given that such information is \emph{not contained}, \emph{codified} or \emph{accounted for} in the standard fundamental description of the situation (given by the complete quantum state and the total Hamiltonian), we conclude that the standard interpretation does not satisfy Bell's criteria, and so it does not offer a satisfactory solution to the measurement problem. It is important to point out, though, that what the above example illustrates is the inability of the standard interpretation to solve the so-called \emph{basis problem} (i.e., the impossibility, given the full quantum description of a system, to pinpoint the appropriate basis to describe actual experimental results); and that solving the basis problem represents a necessary but not a sufficient condition in order to solve the measurement problem. What also needs to be ensured is that the formalism in question is able to accommodate, again, without any external input, the fact that at the end of the experiment what obtains, or at least what we perceive, is only one of the terms of the final state written as a superposition in the appropriate basis (e.g., either one or the other of the terms in equation (\ref{sup})).

How does the situation changes from the perspective of alternatives to the standard formalism, like objective collapse models (e.g., GRW or CSL) or de Broglie-Bohm mechanics? In the first case, given that the spontaneous collapses occur into highly localized states, and that the efficiency of the collapse process increases rapidly with the number of elementary constituents of the system, the theory straightforwardly \emph{predicts} that the macroscopic apparatus will necessarily end-up in a state with well-defined pointer position (i.e., one of the terms in the first basis and not the second); and that is enough to determine, given the quantum description, what it is that the apparatus actually measures. Regarding Bohmian mechanics, a similar thing happens but for a different reason. In such case the apparatus also ends-up in a state with well-defined pointer position, but this time because the fundamental Bohmian description contains, beside the quantum state, the Bohmian particles, which always posses well-defined positions. As a result, the theory also unambiguously dictates that the first basis is the appropriate one to describe the system of our example.

What about CH? Well, the fundamental description of the CH formulation, regarding the situation described above, certainly contains frameworks with final projections into states of the first or the second basis. How does it manage, then, to make predictions? Does it require, as does the standard interpretation, some extra input not given at the fundamental level? Indeed, much of the analysis in \cite{Oko.Sud:14b}, which we review below, aims at showing that this is precisely what occurs for standard measuring scenarios, and that in spite of what proponents of the formalism sometimes claim, CH in fact requires such external input in order to be successful. We conclude, then, that the CH formalism cannot be considered as providing a viable solution to the measurement problem. At any rate, before reviewing in detail our arguments in this respect, we will say a few words about the so-called \emph{second} measurement problem considered in \cite{Gri:15}.
\subsection{The second measurement problem}
An important component of \cite{Gri:15} is committed to argue that CH is preferable with respect to other formulations of quantum mechanics (e.g., objective collapse models, or de Broglie-Bohm mechanics) because, besides solving the standard measurement problem, is unique in solving as well the so-called \emph{second measurement problem}. This second measurement problem is described as the inability of a formalism to explain how an actual experimental result (i.e., an actual pointer direction of an apparatus) is related to the corresponding property of the measured system at a time \emph{before} the measurement took place. The problem, according to the author of \cite{Gri:15}, is that experimental physicists routinely use measurement results in order to infer which properties measured systems possessed before measurements, and so, he claims, a satisfactory quantum formalism must accommodate this common practice. The author also points out that this second measurement problem has received little attention in the literature but that solving it is as important as solving the ``first'' measurement problem.

In spite of these opinions, it seems to us that the reason why almost nobody is concerned with the second measurement problem is because it, unlike the standard measurement problem, does not represent a \emph{conceptual problem} but merely an unfamiliar \emph{feature} present in some formulations of the theory (not unlike the validity of the uncertainty principle or the existence of entanglement). We believe that the author of \cite{Gri:15} finds this aspect to be problematic only because it clashes with our intuitions, common practices and beliefs but that neither the theory nor its usage forces us to assume that experiments must determine preexisting values of measured properties. That is, nothing compels us to ensure that the inference that experimental physicists routinely draw, regarding properties possessed by measured systems, is sound. Therefore, the absence of a ``solution'' to the second measurement problem does not render a theory inconsistent, ambiguous or vague, in contrast to what one faces in the absence of a solution to the standard measurement problem. 

It seems to us, furthermore, that the type of arguments the author uses in order to defend the breakdown of unicity within CH, i.e., arguments to the effect that such feature seems problematic only because it clashes with our imperfect classical intuitions, can well be used in order to defend a theory that does not address the second measurement problem, i.e., a theory incorporating the notion that ``measurements routinely perturb measured systems.'' Such a feature is only strange from the point of view of classical physics, but it does not really represent a conceptual or internal problem for any formalism which contains it. We conclude, then, that the inability of other solutions to the measurement problem to ``solve'' the second measurement problem does not constitute a relevant complication and that the alleged capacity of CH to solve it does not render it superior in any way.
\section{``Measurements according to Consistent Histories'' in a nutshell}
\label{MCH}
As we show in \cite{Oko.Sud:14b}, the application of the CH formalism to measurement scenarios requires the (often implicitly) introduction of input not contained in the formalism at the fundamental level. For example, in order to successfully apply the formalism to a concrete measurement situation, one needs to know in advance (or as Bell would put it, ``using discretion'') what it is that the apparatus one is using actually measures. Only then one is able to choose the appropriate framework, i.e., the one that contains histories for which the apparatus is in a well-defined pointer position when the measurement is completed. 

But what about the Principles of Utility and Equality? Do not they imply that one does not actually need to know in advance what is it that one is measuring? That any choice of framework is as valid as any other, and that one takes the one that, for whatever reason, considers to be more useful? We do not believe that this way of reasoning is valid. That is because, as a matter of fact, given a concrete measurement scenario, it is not the case that all frameworks are equally valid and that one is more useful or informative than the others. The truth is that there is only one framework which contains the history that correctly describes the actual experimental results observed by the experimentalist. And the problem is that CH, in the absence of external input, is incapable of predicting which one is the framework that will do the job.

There are two common responses to our claims. These assert that, for measurement scenarios, frameworks must be chosen either:
\begin{enumerate}
\item To model the experimental situation at hand.
\item According to the questions one is interested in answering.
\end{enumerate}
The problem with the first option is that, as we saw in the example of section \ref{MP}, the fact that a given apparatus actually measures some property is something that cannot be deduced from the CH formalism, even if the full quantum description of the situation at hand is provided. Instead, such knowledge must be discovered by means external to CH (e.g., empirically) in order to be used in the selection of the framework that actually ``models the experimental situation'' at hand. That is, if one is given the description of an apparatus and the subsystem of interest in purely quantum terms, then it is impossible to infer, using only the CH formalism, which is the property that will be actually measured in the laboratory, and so it is impossible, without external input, to select the appropriate framework. We conclude, then, that in order to apply the first recipe one requires input that goes beyond that provided by the theory.

The problem with the second option, i.e., choosing the framework according to the questions one is interested in answering, is that it is not clear what is supposed to be the relation between the framework one chooses, which presumably reflects what one is interested in, and what one in fact observes when the experiment is performed. For example, suppose that we are interested in the spin along $z$ of a particle but that we only have at our disposal an apparatus that measures spin along $x$ (of course, as we just saw, we know what it actually measures through experience, and not because the CH formalism is capable of predicting it). Then, according to the second option above, we must choose a framework that reflects what we are interested in, i.e., a framework with projections onto spin-up and spin-down along $z$. However, it is clear that such a choice of framework will make absolutely no difference regarding what we will in fact observe when we perform the experiment with the equipment provided! 

The choice of framework, then, must be done not according to the questions one is interested in answering, as the second recipe above suggests, but according to the actual experimental set-up. The problem, of course, is that such conclusion takes us back to the first option above which we already saw is unsatisfactory. Therefore, for measurement scenarios, information not provided by the CH formalism is essential in order to select the framework. We conclude that the CH formalism does not meet the second of Bell's criteria described above from which it follows that it does not solve the measurement problem.
\section{Reply to ``Consistent Quantum Measurements''}
\label{Rep}
Section 7 in \cite{Gri:15} directly addresses the criticisms we raised in \cite{Oko.Sud:14b}; in this section we present our response to such a rebuttal. In order to evaluate our objections, the author of \cite{Gri:15} begins by presenting four quotes from \cite{Oko.Sud:14b} (which he labels \textbf{Q1}-\textbf{Q4}) that, he believes (and we agree), summarize our complains. Then he proceeds to comment on them one by one. We will follow a similar procedure.

\textbf{Q1} asserts, in essence, that in order to be useful when dealing with measurements, CH requires the introduction of external input. The author of \cite{Gri:15} accepts that that in fact is the case but defends it by stating the following:
 \begin{quotation}
\noindent [O]ne would anticipate that any plausible fundamental theory of quantum mechanics would contain no reference to [measurements], and therefore additional concepts must be employed in order to describe them. The proper question is not whether elements external to the formalism are used, but instead whether these ``extras'' are needed because one is discussing a measurement and not some other physical process. (\cite[p. 12]{Gri:15})
 \end{quotation}
We find the first sentence above, in the light of Bell quote of section \ref{MP}, very revealing. Let us recall that, according to Bell, a proper solution to the measurement problem contains two aspects: i) \emph{measurements} do not appear at the fundamental level and ii) successful applications do not require external input. Then, if what is said in the quote is true about CH, then it is clear that it does not satisfy Bell's requirements and so it does not solve the measurement problem. As we explained in section \ref{MP}, the point is that it is true that satisfactory solutions should not contain concepts such as \emph{measurement} as part of their fundamental language (as the first part of the first sentence of the quote suggests). But that is not enough; it must also be the case for satisfactory solutions to be capable of describing all situations, including measurements, in terms of the fundamental language of the theory. Therefore, the fact that a formalism requires the introduction of external elements in order to deal with measurements, as CH does, signals that the formalism in question fails to solve the measurement problem. 

The author of \cite{Gri:15} then proceeds to explain in detail how external input (of the kind explicitly not allowed by Bell's criteria) is introduced while using CH for a specific application. All he manages to show, though, is that once such an external input is brought in, the theory manages to deal satisfactorily with the situation. Note however that the same thing would happen if one where using the orthodox interpretation: if one allows the introduction of external input, in the form of knowledge regarding which interactions count as measurements, then the standard quantum mechanical formalism works beautifully. The issue of course is that, as the author of \cite{Gri:15} agrees, the need to introduce this extraneous information renders the orthodox theory problematic. The question is: why does he believe that the situation regarding CH is any different?

The quote \textbf{Q2} starts by claiming that ``[A]mong all the possible frameworks, only one is suitable to describe what in fact we perceive and experience...'' It then continues to stress that CH is incapable of selecting the framework that will do the job. The author of \cite{Gri:15} points out that the topic of perception is subtle and controversial and, taking that as a general remark regarding perceptions, we agree. Nevertheless, the use of the concept of perception we intended should not be controversial or problematic. All we had in mind is the idea that one can meaningfully talk about \emph{actual measurement outcomes}; and it is clear, from many passages in \cite{Gri:15} and elsewhere, that the author of \cite{Gri:15} agrees with that (for example, in \cite{Gri:15} he describes a \emph{measurement} as ``a physical process with some specific macroscopic outcome''). Our point, then, is that CH is incapable of selecting (in advance, and without inappropriate external input) the framework suitable to describe the actual experimental outcomes - a fact that, as we saw, the author of \cite{Gri:15} explicitly agrees with.

\textbf{Q3} elaborates on the idea that ``the fact that a given measuring apparatus actually measures some property is something that cannot be deduced from the CH formalism...'' and concludes that, while using such an apparatus, ``CH is incapable of predicting which framework one must choose.'' As a response, the author of \cite{Gri:15} offers the following:
 \begin{quotation}
\noindent The choice of a framework is one made by the physicist constructing a quantum description, and there is no constraint among the principles of CH that prescribes which one must be used. However, the consistent historian will add that if the ``given measuring apparatus actually measures some property,'' that fact alone is enough to constrain the choice of an appropriate framework..., contrary to the claim in the final sentence in \textbf{Q3}. (\cite[p. 12]{Gri:15})
 \end{quotation}
Once again we wonder how is CH supposed to improve the situation regarding the orthodox interpretation. That is, if one is allowed to use the extra-theoretical knowledge that a ``given measuring apparatus actually measures some property,'' then why would one seek to overcome the standard interpretation?



Finally, quote \textbf{Q4} contains the statement ``the CH formalism is incapable of picking out the right framework.'' As a response, the author of \cite{Gri:15} argues that our use of the notion of the ``right framework'' exhibits a classical prejudice to the effect that ``at any given time there is a single true state of the world.'' He further claims that, at the end of the day, such a prejudice is what lies behind our criticisms of CH. That is, he believes that what truly bothers us about CH is the fact that it does not obey the Principle of Unicity. This, of course, is not correct. While we do find the breakdown of unicity inconvenient, such worry is independent from the objections we present in \cite{Oko.Sud:14b}. The confusion, at least in this particular case, arises because the author of \cite{Gri:15} attaches too much metaphysical baggage to the idea of ``the right framework.'' All we have in mind with such a phrase, as with the idea of ``the framework that correspond with what we perceive'' in \textbf{Q2}, is the following: \emph{the framework suitable to describe the actual experimental outcomes}. 
Given that the notion of actual experimental outcomes is accepted and commonly employed by the author of \cite{Gri:15}, we conclude that his critique of our statements is unfounded. The upshot, once more, is that CH is unable to make correct predictions in the absence of external knowledge.

\section{Conclusions}
\label{C}

We have carefully considered the response offered in \cite{Gri:15} to the criticisms of the CH formalism we presented in \cite{Oko.Sud:14b}; we have found that such response does not manage to refute the basis of our critiques. We do believe, though, that the dispute has served to expose the core of our disagreement with the author of \cite{Gri:15}, as well as with other advocates of the CH approach. In this regard, we have learned that in order to have a sufficiently precise analysis of the way in which the advocates of a scheme such as CH intend to address the measurement problem, one needs a very clear characterization of the problem itself, as well as of what would constitute a satisfactory solution thereof. In order to achieve this clear characterization we have relied upon John Bell's views on the matter and we have used those to pinpoint the shortcomings of the CH approach in dealing with the measurement problem. 
 
The measurement problem in quantum theory has been one of the most highly debated topics in the foundations of physics. Therefore, a critical analysis of the various proposals to address it represents an essential aspect of our search for a better understanding of the workings of nature. The CH approach is one of the most well-developed proposals attempting to resolve the problem, while preserving almost intact the essence of the standard formulation of quantum theory. It must be said that, as such, it has served to highlight the difficulties one must face when attempting such a task. Unfortunately, as we have been able to shown in previous works \cite{Oko.Sud:14a,Oko.Sud:14b} and has been further clarified in this manuscript, the conceptual difficulties that afflict quantum theory are such that they cannot be overcome by a relatively conservative proposal such as CH.

\section*{Acknowledgments}
We acknowledge partial financial support from DGAPA-UNAM project IA400114 (EO) and CONACyT project 220738 (DS).
\bibliographystyle{ieeetr}
\bibliography{biblioCH.bib}
\end{document}